\documentclass[
amsmath,amssymb,
preprint,
]{revtex4-2}

\usepackage{graphicx}%
\usepackage{dcolumn}%
\usepackage{bm}%
\usepackage[utf8]{inputenc}
\usepackage[T1]{fontenc}
\usepackage{mathptmx}
\usepackage{etoolbox}
\usepackage{url}
\usepackage{hyperref}
\usepackage{placeins}
\usepackage{graphicx}
\usepackage{booktabs}

\makeatletter
\def\@email#1#2{%
	\endgroup
	\patchcmd{\titleblock@produce}
	{\frontmatter@RRAPformat}
	{\frontmatter@RRAPformat{\produce@RRAP{*#1\href{mailto:#2}{#2}}}\frontmatter@RRAPformat}
	{}{}
}%
\makeatother


\begin{document}
\preprint{}

\title{Surface and chemical effects on ${^199}$Hg spin polarization relaxation in optically pumped magnetometers}

\author{$^{1*}$Steven K. Lamoreaux}

\date{\today}
\newcommand \hg {$^{199}$Hg\ }

\newcommand \ptrip {6 $^3$P$_1$\ }

\newcommand \psing {6 $^3$P$_0$\ }

\newcommand \rtm {\textregistered~}

\begin{abstract}
Quantum spin magnetometry using optically pumped $^{199}$Hg has been successfully used in many fundamental physics experiments.  A serious problem that has not been resolved is the instability of the $^{199}$Hg spin relaxation rate in atomic vapor cells under irradiation with 254 nm Hg resonance light. In this paper, previously obtained data are re-analyzed or analyzed for the first time. The effects of impurities of H$_2$ and O$_2$ are elucidated, and possible ways to stabilize cells are discussed. Surface states originating from the an der Waals interaction of \hg with fused silica are analyzed and shown to be critical to understanding relaxation mechanisms. A discussion of the possible use of a mixture of N$_2$O and other gases is presented.
\end{abstract}
\affiliation{$^{1}$Physics Department, Sloane Physics and Wright Laboratories, Yale University, P.O.Box 208120, New Haven, CT 06520-8120 USA}

\email{Steve.Lamoreaux@yale.edu}

\maketitle

\section{Introduction}\label{sec1}

This research article is different from most because it describes work using optically pumped \hg that has been done over the last 40 years, some of which was never published, and with only partial theoretical analysis.  As such, a review article is not possible due to the limited specific literature to review.  Typically, studies of cell wall-induced atomic polarization relaxation times, defined here as $\tau_w$ (with wall relaxation rate $\Gamma_w=\tau^{-1}_w$), were performed in the context of building an apparatus to measure some fundamental process of greater immediate interest; as typical in such work, tedious technical details are barely left as a footnote if presented at all.

In this research article, optically pumped \hg will be put in context with other work in the field. In particular, the importance of surface states and the van der Waals interaction of atoms with walls will be discussed together with a theoretical analysis.  

The study of the optical pumping of ground state atoms had been expanding through the 1950s and Hg was among the very atoms first to achieve $\tau_w$ of 100's of seconds. One of the most important developments in the field is transmission monitoring of the ensemble spin polarization, an idea due to Dehmelt \cite{dehmelt} who patented a magnetometer design that is still in use.\cite{dehmeltp} (See \cite{hap,budker} for reviews of the field.) 

The optical pumping of the ground states of $^{199}$Hg and $^{201}$Hg was first achieved in the early 1960s by B. Cagnac at the Laboratoire de Spectroscopie Hertzienne de l'ENS (now named the Kastler-Brossel Laboratory).\cite{cag} The theory was fully developed and tested by Cohen-Tannoudji in his Ph.D. studies\cite{ct1,ct1a}, and he has written a general review of the early work at this laboratory. \cite{ct3}  

Hg is a $^1$S$_0$ atom, so the ground-state magnetic moment, if any, is due only to the nucleus; for the two isotopes with significant natural abundance, only two have non-zero nuclear spin: 199 $I=1/2$, and for 201, $I=3/2$.  In addition to applications to nuclear magnetic resonance-based gyroscopes \cite{ussrgyro,singer}, optically pumped Hg has found minimal practical applications because alkali atoms, with an unpaired electron spin and hence larger atomic gyromagnetic ratios, can provide much higher sensitivities in magnetometer applications.  Optically pumped Hg has, however, found application in several fundamental experiments. 

In the early 1980s at the University of Washington, Seattle, Prof. E.N. Fortson identified optically pumped Hg as useful for a number of fundamental experiments, the first to search for spatial anisotropy that used the mass quadupole moment of the $^{201}$Hg nuclear spin to limit the coupling to its orientation or motion through absolute space in a modernized Hughes-Drever experiment.\cite{lamoreaux_hd,lamoreauxphd}.  The optical pumping cells developed for this work were also used to test a proposed non-linear extension to quantum mechanics, which incidentally provided a gyroscopic measurement of the Earth's rotation.\cite{venema}

These experiments used optical pumping cells that were heated to around 400 $^\circ$C and used the $^{199}$Hg nuclear spin as a comagnetometer (a term we invented, using ``co" from Latin, meaning ``with" or ``together") to eliminate frequency changes due to magnetic field fluctuations that would interfere with the fundamental physics measurements via $^{201}$Hg.  

These experiments were side projects; the main research goal at the time (1983-90) was to develop an optical pumping cell in which an electric field could be generated inside the cell volume and to determine whether the applied electric field leads to a change in the Larmor precession frequency of the nuclear spin of $^{199}$Hg atoms.  If the atom has an electric dipole moment in addition to the magnetic one, there will be a shift in the Larmor frequency and this will imply the existence of a fundamental time-reversal violating interaction(s) with the $^{199}$ Hg atom, the effects of which are expected to be enhanced in high atomic number atoms. (See \cite{lamkhrip} for a review of the field, and a more recent review by Chupp et al. \cite{chupp}.)  

The essential problem was to find cell coating materials that give a long $\tau_w$ for \hg and also allow the application of high voltage, without the need to heat the cell to high temperature.  Inclusion of a relatively inert ``buffer gas together with low (room) temperatures is required for high voltage stability.  An earlier experiment in the Fortson group employed spin-exchange polarized $^{129}$Xe,\cite{vold} for an EDM experiment, and $\tau_w$ near room temperature on the order of a thousand or more seconds were routinely achieved in the presence of a field on the order of a few kV/cm.  As \hg offers a higher sensitivity to possible fundamental interactions of interest in addition to an intrinsically better measurement sensitivity, a vigorous research effort was launched to establish an \hg EDM experiment.  

Attempts to find another spin-1/2  atom to serve as a comagnetometer in the same cell as \hg were unsuccessful; however, a workable cell was developed with \hg alone, although $\tau_w$ would degrade with exposure to 254 nm Hg resonance radiation. ($^{201}$Hg has a very short $\tau_w$ in room-temperature cells due to its nuclear quadrupole moment, which in addition, anyway, precludes its use in EDM experiments.) This problem persists to this day in that the construction of stable cells remains a hit-and-miss.  

Optically pumped $^{199}$Hg has found use in several other fundamental experiments, for example, a tests for axions and their long-range interaction from both spin-polarized and unpolarized material bodies.\cite{hunter1,hunter2} 
All of these experiments, including the $^{199}$Hg EDM experiments, employ closed cells in that they are filled with appropriate amounts of $^{199}$Hg vapor and other gases and permanently sealed off from a preparation manifold.

In addition, $^{199}$Hg has served as a comagnetometer for several neutron EDM experiments, the first application being at the Institut Laue-Langevin.\cite{lamoreauxnim,ksmite} In these experiments, the storage cell for ultra-cold neutrons (UCN) and $^{199}$ Hg together needed to be developed, and $\tau_w$ stability problems persisted in these experiments. A coating that is suitable for both the atomic spin and storage of UCN was found, deuterated polystrerne (DPS) \cite{dps}, and to date there does not appear to be a fully satisfactory replacement. These experiments are ``open" in that background gases (including $^{199}$Hg) can be pumped out of the cell.  The techniques were greatly refined by the Paul-Scherrer Institute, \cite{psi,psiedm} leading to the most sensitive neutron EDM measurement performed so far. This apparatus was also used to search for axion coupling between a Cu mass and nuclear spins.\cite{psiaxion}

Better wall coatings for \hg with longer and stable ensemble nuclear spin polarization relaxation times $\tau_w$ are of interest to several ongoing experiments, including the neutron EDM experiments at the Paul-Scherrer Institut \cite{psiedm2} and Los Alamos \cite{lanledm}, and tests of the equivalence principle by Hunter et al. \cite{hunter3} which is the principal motivation for the study presented here.  

The $\tau_w$ issues for closed and open cells are related, as will be examined here.   
The purpose of this paper is to briefly review the development of $^{199}$Hg optical pumping cells for EDM and other fundamental experiments and to provide an analysis of data that has not been published.  This data sheds light on the chemical and physical mechanisms of the $\tau_w$ instability, which, when elucidated, will help identify what is needed to achieve reproducibility in cell fabrication.

As an aside, although the limit of the \hg EDM \cite{uwfinedm} is numerically much lower than the neutron EDM
(this allows it to be used as a comagnetometer), theoretical interpretation of the limits of the \hg EDM results in a loss of sensitivity to fundamental interactions, so the impacts of both types of experiments on testing the standard model of electroweak interactions are similar. EDM experiments based on optically pumped \hg alone have potential for improvement, although the lack of a comagnetometer within the cell volume remains a potential ultimate limitation.

\subsection{Development of \hg EDM cells}

The basic optical pumping technique used in the development of \hg cells employed circularly polarized resonance radiation and transmission monitoring \cite{dehmelt} of the ensemble spin polarization.
Resonance light from a $^{204}$Hg microwave discharge lamp contains a single 254 nm line (corresponding to the $6^1 S_0$ to $6^3P_1$ transition) because there is no hyperfine structure for this isotope.\cite{cag,ct1a}   The single spectral line overlaps the $F=1/2$ ground state to $F=1/2$ excited state levels of $^{199}$Hg.  If a cell containing atoms is placed in a few mG magnetic field oriented along the light propagation direction, the ground state atoms will be excited only from the $m_F=-1/2$ if the light has left-handed circular polarization (which carries $+1$ unit of angular momentum).  The decay back to the ground state can go to either level; however, atoms accumulate in the $m_F=1/2$ level because these never go through an optical pumping cycle.  Eventually, the atomic nuclear spin polarization reaches equilibrium, and the transparency of the cell increases. The equilibrium spin polarization is, when the only spin polarization relaxation is due to wall interactions, is
\begin{equation}
	P=\frac{\Gamma_p}{\Gamma_p+\Gamma_w}
\end{equation}
where $\Gamma_p$ is the optical pumping rate or the inverse of the pumping cycle time, $1/\tau_p$, and $\Gamma_w=1/\tau_w$ is the wall relaxation rate.  For most of the subsequent discussion, it will be assumed that wall relaxation is dominant over magnetic gradient effects or interactions with other atoms or molecules in the gas phase.  In the weak field limit ($\ll 100$ G), in the absence of magnetic gradients, the longitudinal and transverse (coherence) relaxation times are equal ($T_1$ and $T_2$ in nuclear magnetic resonance terminology). (See   \cite{hap} for a comprehensive review of ground-state optical pumping, in which Fig. 1 is a pictorial of this specific case.)

If the magnetic field is suddenly switched to be perpendicular to the light (in time much faster than the inverse Larmor frequency), $P$ becomes time dependent, and the transmitted light is modulated at the Larmor frequency, which is about 7 Hz for a 10 mG magnetic field. The modulation level will exponentially decay with a total rate given by the sum of $\Gamma_p$ and $\Gamma_w$, so to determine $\tau_w$, the total rate must be extrapolated to $\Gamma_p=0$. 

For discharge lamps, the non-resonant background light was typically only 20\% of the emitted light, with the dominant noise source being the shot noise in the detected photons.  Recent experiments have employed frequency-quadrupled semiconductor lasers, which have reduced non-resonant light and higher-efficiency detectors. Furthermore, such lasers are tunable, which allows for a higher density of \hg because the laser can be tuned away from the absorption line center. 

Research toward developing a room temperature $^{199}$Hg optical pumping cell was begun in 1984, with the first trials using fused-silica blown bulbs of about 18 mm ID.  The kinetic mean free path (mfp) between wall collisions is given by
\begin{equation}\label{mfp}
	\ell=\frac{4V}{A}=\frac{4}{3}R
\end{equation}
where $V$ and $A$ are the internal volume and surface area, and $R$ is the radius of the sphere.  These cells also had fill stems about 20 mm long where they were sealed from the vacuum system after being filled with natural Hg 0 $^\circ$C vapor pressure. 

The first cell had the inner surface coated with paraffin (dotriacontane) ``wax" and had a $\tau_w$ of about 40 s.  For a larger cell, as envisioned for EDM work, with a factor of 2 greater mfp, assuming homogeneous surface relaxation, $\tau_w$ would be 80 s and good enough for an EDM experiment.

Under illumination with the 254 nm pumping light, the cell's $\tau_w$ deteriorated. It could be temporarily restored by remelting the wax; this potentially ominous effect was not fully appreciated at the time.

The next batch of spherical cells had siliconizing agents applied to the inner surface, both Surfasil\rtm and Aquasil\rtm were tested, both with and without wax, and an Aquasil\rtm cell was filled with 40 torr N$_2$ buffer gas (Linde\rtm spectroscopic grade).  The wax was introduced only after the cells without wax were removed from the vacuum system. It was observed that the wax did not wet the siliconized surfaces but accumulated in individual droplets, so the wax was driven into the cell stem, where it formed a sort of crystalline plug. 

All of these cells had a $\tau_w$ of only about 15 s.  This was thought to be due to the damage to the coating that occurred with the cells being sealed off, the poor character of the wax plug, or magnetic field gradients in the case of the buffer gas cell (the magnetic shield was very crude), which had a 5 s $\tau_w$.  The Aquasil\rtm coated cell (10\% dilution in acetone, followed by two rinses with acetone and drying by blowing air into the bulb) had the best long-term performance.  Warming the cell to 120$^\circ$ C increased useful life to 60 s, with a modest permanent improvement at room temperature. 

Emboldened by these results, we made cells similar to those used in the $^{129}$Xe experiment \cite{vold} but with Hg and only N$_2$ as a buffer gas required for high-voltage stability.  These cells had a fused silica cylinder of 25 mm OD, 23 mm ID, 10 mm high with a fill stem that enters radially onto the side of the cylinder, and the cylinder was capped with fused silica disks of approximately 38 mm diameter, glued with Varian Torr-Seal\rtm epoxy.  To provide a means of high voltage (HV) application within the cell to establish an electric field, a variety of electrode materials were tried; however, exactly the same electrode coating material (tin oxide) as used in the $^{129}$Xe experiment, coated with wax, and with Aquasil\rtm on the cylinder body gave excellent performance.  

\begin{figure}[!h]
	\centering
	\includegraphics[width=120mm]{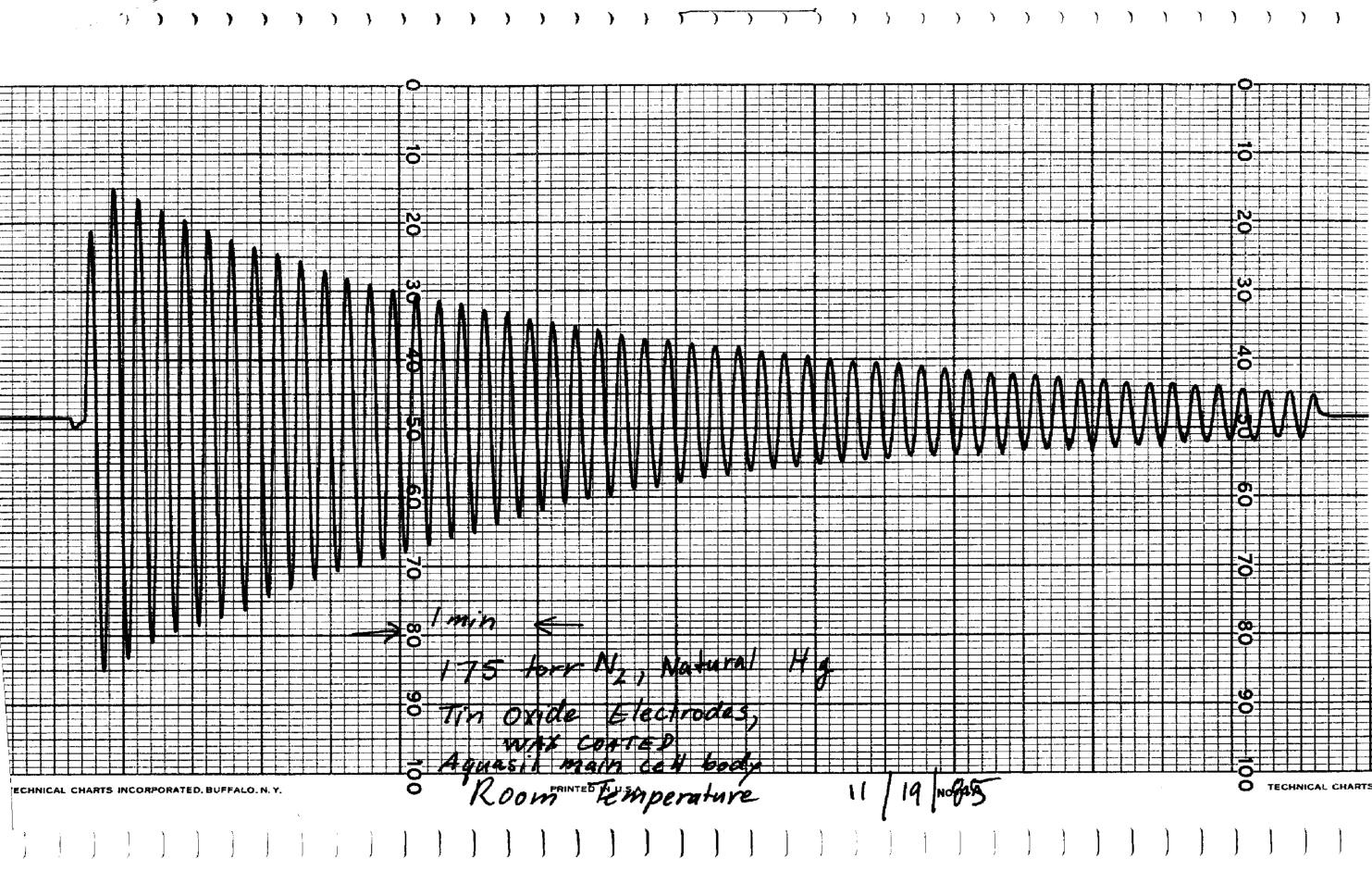}
	\caption {1st successful EDM cell, probe light limited the spin polarization relaxation time to 200 s.}
	\label{200sec}
\end{figure}

The data shown in Fig. \ref{200sec} was obtained by optically pumping the system with circularly polarized resonance light propagating along the direction of a weak magnetic field (10 mG), which was suddenly switched off with the simultaneous sudden application of a second field (10 mG) perpendicular to the light, as described earlier. 
The angular momentum states for the transition in $^{199}$Hg are optimal for optical pumping because the sample becomes transparent (for pure $^{199}$Hg) when fully polarized, and the modulation on the light during free precession can be nearly 100\% for a fully polarized ensemble; however, the modulation is reduced to about 15\% with a natural isotopic mixture of Hg, as was used throughout the cell development period and for the first \hg EDM experiment. The actual Larmor frequency was about 10 Hz, which was mixed with the signal from a constant oscillator to bring the precession frequency down to where it could be measured with a chart recorder.

At the minimum controllable light level that we could achieve with our system, $\tau_w$ was 300 s, and met the 100 s goal for cell development (because $1/f$ noise magnetic field noise required a relatively fast electric field reversal rate, so a longer time was unnecessary). A quick test in the dark indicated a lifetime in excess of 1000 s, which is consistent with other $^1S_0$ nuclear-spin 1/2 atoms, particularly $^{129}$Xe.  We therefore thought that we were done with cell development, as cells made in this manner proved to be stable under the application of high voltage.

Unfortunately, our hopes were dashed when we discovered that the cells were unstable against the loss of Hg. The first version of the \hg experiment was based on an atomic oscillator configuration, where the static magnetic field (quantization axis) is oriented 45$^\circ$ away from the light propagation direction, and an oscillating field close to the Larmor frequency is applied. The atomic system develops a coherent polarization oscillation, with the precession evident in the transmitted light modulation.  The phase shift between the atomic precession and the oscillating field is proportional to the degree of detuning between the Larmor and oscillating field frequencies and can be measured with a phase-sensitive detector.

Two such cells were operated together with electric fields directed oppositely, with the sum of the frequency difference used to stabilize the average applied field, and the difference used as a correction signal to stabilize the gradient field between the cells. The gradient correction signal was sensitive to an EDM because the electric field in the cells was directed oppositely.\cite{lamoreauxedmprl}

\begin{figure}[!h]
	\centering
	\includegraphics[width=80mm]{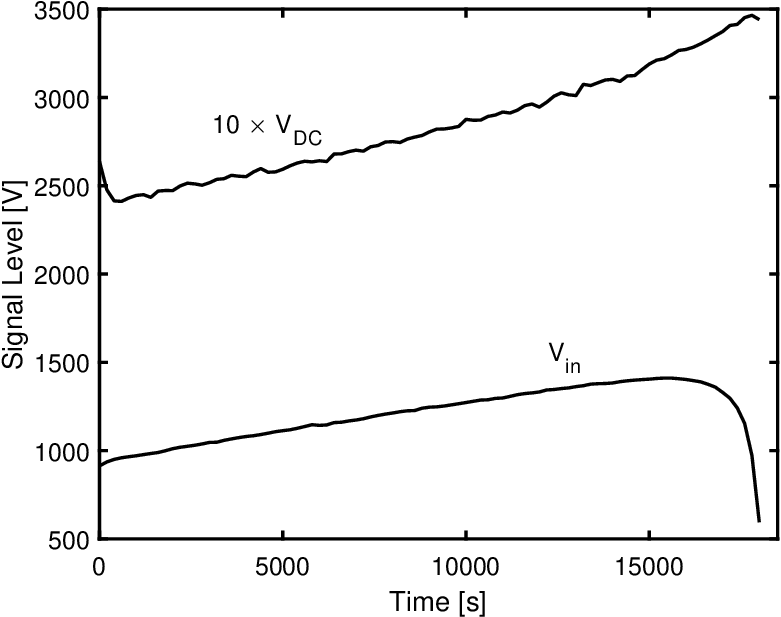}
	\caption {$V_{DC}$ is a measure of the transmitted light intensity (multiplied here by 10 to offset the plot from $V_{in}$), and its increase indicates a general upward drift of the light intensity, with a rapid loss of Hg atoms near the end time, as further evidence by the loss of the oscillator in-phase signal $V_{in}$. This data was manually digitized from plots in \cite{lamoreauxphd}, Fig. 8.2, cell 2.}
	\label{evolution}
\end{figure}

The in-phase oscillating transmitted light signal provides a measure of the sample atomic polarization, times the atomic density.  Shown in Fig. \ref{evolution} is one of the data runs in which the loss of Hg in the cells is evident at the end of the run when the oscillator signal $V_{in}$ fades.

This data has some interesting features that can be understood by considering the degree of polarization of the ensemble,
\begin{equation}
	P=\frac{\Gamma_p}{\Gamma_p+\Gamma_w+\omega_1}
\end{equation}
where $\omega_1$ is the oscillating field amplitude.
Normally, the light intensity in our experimental work was very stable, so this data run is slightly anomalous; however, given the haste with which we had to set up and test the system, it is understandable.
At the start of the experiment, the light intensity and $\omega_1$ were adjusted such that $\Gamma_p\approx \omega_1$ and $\Gamma_p+\omega_1 \approx 50\ {\rm s}^{-1}$,
with approximately 30\% precision.  Noting that the light intensity is proportional to $\Gamma_p$, and that $\Gamma_w=\tau_w^{-1}\ll \Gamma_p$, the in-phase signal should scale with the light intensity as its product with $P$:
\begin{equation}
	V_{in}\propto \frac{\Gamma_p^2}{\Gamma_p+\omega_1}\approx \frac{\Gamma_p}{2}.
\end{equation}
This scaling works up to about 13000 seconds, at which time $V_{DC}$ begins to increase rapidly, while $V_{in}$ begins to fall.

This marks the onset of the loss of Hg due to reactions with atomic oxygen, or more accurately, ozone O$_3$ which takes some time to build up. This process will be described in some detail later in this article;  is a two-step process by which Hg$^*$ reacts with O$_2$ to form O$_3$ and atomic O.   O$_3$ reacts irreversibly with Hg to form HgO, resulting in an increase in transmitted light.  The initial O$_3$ concentration is zero, so the loss of Hg does not begin until it is formed, with the loss rate increasing over irradiation time.
The differential equation governing this has a loss rate that increases linearly with time (neglecting the loss of Hg so this is approximate, for the early time development),
\begin{equation}
	\frac{d\rho}{dt}=-\alpha t\rho
\end{equation}
where $\rho$ is the Hg density, and $\alpha$ is a constant rate parameter that is proportional to the resonance light intensity.  A full model with both varying Hg density and light intensity gives a good representation of the observed time dependence. 
In the case of constant intensity, the solution is $\rho(t)=\rho(0)e^{-\alpha t^2}$ which explains the sudden drop in $V_{in}$ at the end of the run. If a cell had been irradiated, the density would continue to change as $\rho(t)=\rho(t_0)e^{-\alpha t_0 t}$, assuming the ozone concentration is constant.

It can be further noted that the change in $V_{DC}$ when the Hg disappears, accounting for the upward drift in light intensity, implies that the Hg light absorption at the beginning of the run was only about 15\%. This is most likely due to an earlier loss of Hg from the same photosynthesized reaction.  Even if a cell is briefly illuminated, there will be some ozone formation, and although at low concentrations the rate of Hg loss is slow, nonetheless the reactions will continue until the ozone is absorbed or reacted with the materials within the cell, as described above. Note that these cells were operated in the strong pumping regime, so it is not possible to deduce if there was a change in $\tau_w$ based on these data.  

The loss of Hg limited the duration of the EDM measurements for each cell pair, which for the case discussed was about 5 h.   
In obtaining the first \hg EDM limit, we opted to manufacture new cells (in batches of eight) that would be used in pairs until their Hg density decreased to a point where the oscillators failed. The net result of this imperfect but successful study was published.\cite{lamoreauxphd,lamoreauxedmprl}

In principle, a source of Hg could have been incorporated into cells and thus keep the Hg density constant; however, the optical density would have been too high for any practical system at the time, as the proper natural Hg density roughly corresponds to the $0^\circ$ vapor pressure. (\hg accounts for only 15\% of the total absorption of light produced by $^{204}$Hg; later cells employed enriched \hg.)  At the time, no tunable sources of resonance light were available. This allows probing away from the line center, as is done currently using quadrupled semiconductor lasers.

\begin{table}[!h]
	\caption{Excitation and bond energies of interest. Most are available in chemistry textbooks (e.g., \cite{lt}); citations are provided for the rarely encountered bonds.}
	\label{Parameter10}
	\centering
		\begin{tabular}{lll}
			\toprule
			State & Energy & Comment/Reference\\
			or Bond & (enthalpy) $[e{\rm V}]$&\\
			\midrule
			$^3$P$_1$ Hg$^*$ &4.89 &allowed transition (oscillator strength $f\approx 1$) \\
			$^3$P$_0$ Hg$^*$  &4.55 & metastable, populated by N$_2$ collisions with $^3$P$_1$ Hg$^*$\\
			H-H & 4.45 &\\
			O-H & 4.76&\\
			(H:O)-H&4.8\\
			C-H&4.26&\\
			(H$_5$C$_6$)-H& 4.46&extra binding energy compared to C-H due to resonance\\
			&& stabilization of the phenyl ring\\
			F-H& 5.80&\\
			Hg-H& 0.37 & the extra electron in HgH destabilizes the bond\\
			(H:Hg)-H& 3.0 & thermodynamically unstable \cite{hgh2} \\
			O-Hg& 0.94&\\
			C-Hg& 0.72& \cite{ref1}\\
			(O:O):O&1.5&\\
			C-F&4.52&\\
			C-C&3.58&\\
			C$=$O& 8.23&\\
			C$=$C&8.62&\\
			O$=$O&5.12&\\
			(O:C)-H& 1.64& formyl (free) radical,\\&& 0.087 $e$V activation energy to form formaldehyde\cite{coh}\\
			(HCO)-H& 3.82& formaldehyde\\
			C$\equiv$O&11.04&\\
			N$\equiv$N& 9.74 &\\
			(N$_2$):O &-0.85 & exothermic, 1.7 $e$V activation energy
			(can detonate) \cite{n2oae};\\
			&&scavenges O$_3$ \cite{n2odep}\\
			
			\bottomrule
		\end{tabular}
\end{table}

\subsection{Oxygen}

The next stage of experimental development was to identify the Hg loss mechanism. A chapter in \cite{herzberg} describes photosensitized reactions involving \ptrip excited-state Hg atoms. Specifically, excited Hg$^*$ 
in collisions with O$_2$ transfers most of its energy to the molecule, which subsequently reacts with another O$_2$ to form O$_3$ (ozone) and atomic oxygen O.  O$_3$ subsequently reacts with Hg to form HgO and O$_2$, thus losing an Hg atom from the gas. There is no isotopic enrichment in the HgO formed in this process, which led to the full elucidation of this process \cite{ozone1} that has been known for close to 100 years.\cite{ozone2} It can be understood that such a series of reactions is possible from the bond energies given in Table 1. 

Armed with this knowledge, we took steps to remove oxygen from the nitrogen buffer gas by bubbling the gas through liquid amalgams of Mg-Hg or Al-Hg followed by a liquid nitrogen trap to remove Hg vapor.\cite{jpjphd}  After such purification, the Hg density became constant; however, the cell's $\tau_w$ rapidly degraded with resonance light irradiation.

A further literature study of photosensitized reactions indicated that metastable \psing Hg atoms can be readily formed with collisions between \ptrip Hg atoms and nitrogen molecules, due to resonance with the rotational states of N$_2$. The \psing excited atoms still carry considerable energy, see Table 1, and can damage the wall coatings. 
The metastable atomic state lifetime has been estimated as a few seconds, so there is ample time for diffusion to the cell wall.

The destructive nature of \ptrip and \psing atoms has been known for nearly 100 years (this was known before the possibility of producing ozone \cite{cf}).  As shown in Table 1, these excited states have enough energy to break most single bonds.  

Indeed, even bare fused silica is prone to damage by these excited atoms.  In our earlier work with testing spatial isotropy, we discovered that fused silica optical pumping cells that were irradiated while at room temperature would suffer irreversible damage as manifested by a shorted $\tau_w$. The damaging effects on silica have been known since 1930 \cite{silica} but apparently not by us.  

There was a vague notion that hydrogen was responsible for the degradation of the $\tau_w$; however, a model was never developed and the notion tended to be dismissed due to the successful operation of hydrogen masers (more on this later). 

The best option to quench \ptrip (that is, perturb the excited state causing a radiative transition or energy transfer)  directly to the ground state and also to quench \psing (which is not quenched by nitrogen) is carbon monoxide, CO.  Cells were made with 5\% CO and 95\% N$_2$ and the useful life expectancy of cells under UV irratiation improved.  The University of Washington \hg EDM experiments over the next decade employed such cells.

\subsection{Further Advances}

The latest \hg EDM experiment is reported in \cite{uwfinedm}, and incorporated several improvements. First, it used a pulsed pump and probe method that gave a factor of nearly 2 in the average atomic polarization. The spin precession was detected by use of the Faraday effect (linear polarization rotation) of second light beam that was not resonant with any \hg 254 nm transitions. Thus, there was little or no optical pumping relaxation due to the detection light, and the pump light was blocked during a measurement.  (It should be noted that a similar system is described in \cite{ussrgyro}.)

This high performance system was made possible by advent of 254 nm tunable lasers, and the tightly focused beams allowed the use of higher quantum efficiency small area photodiodes.  The small diameter laser beams also meant that the cell walls were not fully bathed in resonance light, and exposure to Hg$^*$ was reduced by the use of non-resonant detection light.  A higher \hg density was also possible when using the Faraday effect and lasers, which could be tuned away from the line center.
The cells contained pure CO as a buffer gas, instead of a mixture with N$_2$. Obviously, care was taken to remove oxygen; however, the level of hydrogen contamination is not reported.  

These cells were carefully constructed with polished, optically flat end disks and cylinder edges, where a vacuum adhesive was applied.  Instead of Varian Torr-Seal epoxy, Space Environment Labs Vacseal silicone-resin-based leak sealant was used.  This material has a much lower outgassing rate compared to Torr-Seal, which, together with the close fit between the cylinder and the end-cap disks, minimized the exposure of contaminants that might be generated by the sealant.  The cell components were carefully cleaned by refluxing in hydrochloric acid.  These cells had a stable $\tau_w$ of at least 100 s.  This is notably lower than the 300 s light-limited spin polarization relaxation time obtained in the first EDM cells which had a small amount of O$_2$ gas. Although the presence of O$_2$ resulted in a loss in Hg (gas), it apparently improved the spin relaxation properties of the surface.  It is well known that some cured silicone resins are slightly permeable to oxygen, \cite{oxyperm} which could be a reason for the stabilization of the cell spin polarization relaxation time.  
\cite{lindpc}

Four cells were used in this experiment, and later attempts to reproduce these results have not been entirely successful.

\subsection{Hydrogen}

Since CO was identified as a useful buffer gas, there has been a serious suspicion that the cell degradation is due to hydrogen.  However, this notion was often met with skepticism, with a counterexample being the Hydrogen Maser: If H atoms were a destructive agent, how could a maser ever operate?  In fact, this atomic H is a problem for masers.  

One of the first successful maser bulb coatings was Teflon\rtm using DuPont TFK Clear Finish 852-201, while in later work it was applied using
DuPont FKP Teflon product code 120. Quote \cite{teflonwater}   
\begin{quote}
	Although the wall coating procedure described here is quite reliable, the surfaces are not entirely inert, and it is possible that this is due to a contaminant in Teflon. New surfaces are currently being investigated.
\end{quote}
In the same year, Berg, one of the coauthors of this work, published a single-author follow-up paper stating that excess hydrogen from water-based Teflon coatings and non-UHV compatible vacuum components (rubber o rings, etc.) should be eliminated and had empirical evidence to support that notion.\cite{berg}

Later work shows that atomic H actually visibly degrades Teflon coatings, as evidenced by the damage at the atomic beam focus point, on the surface opposite the entrance aperture of the maser bulb.\cite{vessot} 

We encountered similar problems with \hg when we used a water-based Teflon coating; coatings made from pure PTFE (polytetrafluoroethylene) in a volatile non-polar carrier give stable and reproducible $\tau_w$; however, the $\tau_w$ are limited to 15-20 s per cm mfp.

\subsubsection{Autocatalytic Hydrogen Production}

With the inclusion of CO in the cells, we had hoped that excess hydrogen would form formaldehyde and get absorbed into the wax.  As can be seen for the bond energies in Table 1, essentially all materials are susceptible to attack by excited Hg atoms or by atomic H, so the notion of sequestering hydrogen in volatile molecules is not useful.

This leads us to surmise that once hydrogen is present in the system, the rate of generation of more hydrogen actually increases, in the presence of exited-state Hg. This is because the two H atoms from the splitting of each molecule split by an interaction with an Hg$^*$ will more likely combine with and remove H atoms from a wall coating molecule and form new H$_2$ molecules, rather than finding and recombining with their former partners.  Thus, the degradation process accelerates and, ultimately, cells that initially could be rejuvenated by ritualistic treatments that involve melting wax, eventually had $\tau_w$ degraded to the point that the cells become useless.  

Furthermore, atomic H itself is a free radical, and its presence in the cell along with the buffer gas can cause the relaxation of spin polarization of \hg, which adds to that due to the walls.  The spin polarization relaxation time resulting from this is estimated in the discussion section.

\subsection{Surface States}

A previous study, with its carefully acquired large data set, \cite{romalis}, has been very helpful in understanding the dynamics of spin relaxation of \hg.
Here we add to the analysis presented there and recapitulate and expand the analysis presented in \cite{mysurface}. 

First, consider Eq. (3) of ~\cite{romalis}. This equation is true if there is a single bound state for an \hg atom on the cell surface.  However, the number of bound surface states grows with the mass of the bound atom,\cite{jmp}, a point that has rarely, if ever, been taken into account in previous similar work (see, e.g., \cite{xesurf}). 

The number of states and their energies can be estimated using the Wentzel-Kramers-Brillouin (WKB) approximation, which is an approximate method for determining the bound state energies and the wavefunctions of a massive particle bound in a spatially varying potential\cite{wkb}.  The van der Waals potential (short-range Casimir-Polder force) between an \hg atom and a dielectric surface is given by
\begin{equation}\label{vdw}
	V(z)\approx\begin{cases} \frac{C}{(d_0 +z)^3}& z\geq 0\\
		\infty & z<0 \end{cases}
\end{equation}
where $C\approx -6\times 10^{-4} \ e$V nm$^3$ for a heavy polarizable atom near a dielectric surface, with $z$ the distance from the surface in nm and $d_0\approx 0.14$ nm representing the sum of radii of the Hg atom and atoms bound in the fused silica surface.  The coefficient can be calculated using the formalism in ~\cite{vdw95} with the dielectric function for fused SiO$_2$ in ~\cite{cs}, Fig. 4, where the calculation was done for Cs. The value for Hg is about 1/6 of that for Cs, due to the shorter wavelength of the allowed electric dipole moment transition of Hg, 180 nm compared to 895 nm and 854 nm (D1 and D2 transitions) for Cs, together with the reduced dielectric response of SiO$_2$ at shorter wavelengths. This leads to a reduction in the overlap integral. 

\begin{figure}[!h]
	\centering
	\includegraphics[width=80mm]{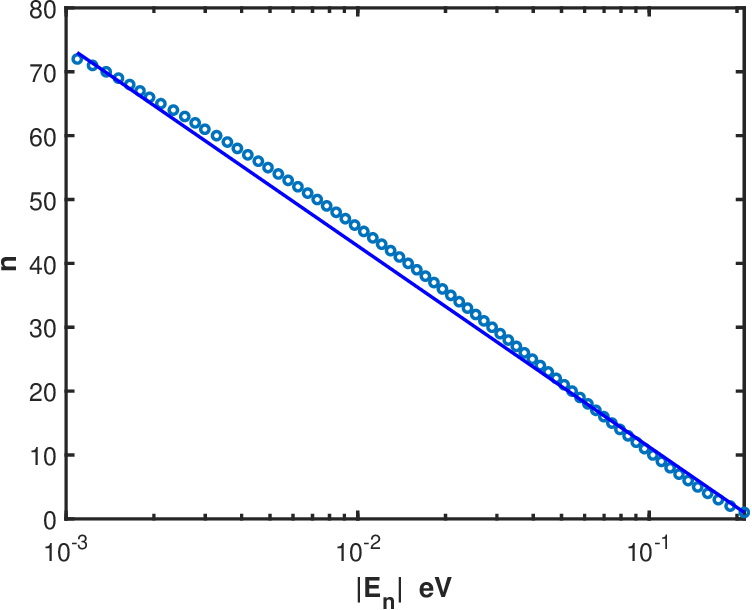}
	\caption {Bound state energies as a function of wavefunction nodes plus 1, with a functional approximation to the cumulative distribution function for the states shown by the solid line. }
	\label{nede}
\end{figure}

The number of bound states is approximately the number of nodes for the least tightly bound WKB solution, which is $n_{max}\approx 72 $ for $m=199$, and the two lowest states are approximately -0.18 $e$V and -0.21 $e$V. Here, a plot of $E_n$ versus $n$, the number of wavefunction nodes plus one, is shown in Fig. \ref{nede} .  
The density of states is a nearly continuum; also plotted is 
\begin{equation}
	F(E_n)=\frac{N_{max}}{\log(E_{max}/E_{min})}\log(E/E_{max})+1
\end{equation}
which is a good approximation to a cumulative distribution function of the bound states.  

Therefore, Eq. (3) of \cite{romalis} needs to be modified, and one possibility is to instead take the average sticking time as
\begin{equation}
	\tau_s\propto\sum_{n=1}^N \tau_{0n}e^{|E_{an}|/kT}
\end{equation}
where we allow $\tau_0$ to depend on $n$, and $E_{n}$ are the bound state energies determined by the WKB approximation. However, since each level is likely to have a different sticking or residence time, a more profitable approach is to calculate the fraction of atoms on the surface by using the partition function.  
The binding energies of the levels are occupied according to the Boltzmann distribution, so we can determine the average fraction of the total number of atoms distributed among the bound states and scale that according to the time between wall collisions.

Assume that the atoms move unhindered while in the main cell volume (away from the wall) and at the wall there are bound states with energy $-E_n$ (for the following, we take $E_n=|E_n|$).  In a bound-surface state, we can assume that an atom moves more or less freely parallel to the surface. The model generally assumed for this problem is that the atom hops between lightly bound regions.\cite{hap}  For simplicity, assume a spherical cell with volume and area $V$ and $A$, $\beta=1 / k_b T$, and $p$ is the momentum of the atom. Using simple statistical theory, the probability $\epsilon_n$ that an atom spends time in the surface state $n$ is (assuming that this probability is dynamic and very small),
\begin{eqnarray}
	\epsilon_n&=&\frac{A \hbar^{-2} \int_0^{\sqrt{2 m E_n}} \exp \left[\left(E_n-p^2 / 2 m\right) \beta\right] d^2 p}{V \hbar^{-3} \int_0^{\infty} \exp \left(-\beta p^2 / 2 m\right) d^3 p} \\
	&\approx& \frac{\pi A e^{\beta E_n} \lambda_T^{-2}}{V \lambda_T^{-3}} = 3 \pi e^{\beta E_n} \frac{\lambda_{T}}{R}
\end{eqnarray}
where $\lambda_T=\hbar(2 \pi \beta / m)^{1 / 2}$ is the thermal wavelength, $\beta=1/k_bT$,  $m$ is the mass of an atom, and $R$ the radius of the cell, about 1 cm. Approximately, for Hg at $293 \mathrm{~K}$, $\lambda_T=$ $7 \times 10^{-10} \mathrm{~cm}$. For the most strongly bound WKB state,  $E_n\approx0.21\ e$V,  and with $R=1 \mathrm{~cm}$ determines $\epsilon\approx 3 \times 10^{-5}$.

The total fraction of (temporarily) bound atoms on the surface is given by the sum of $\epsilon_n$ for all states.  This could be done by summation; however, an approximate analytic formula better illustrates the effects of temperature variation. The bound states as determined by the WKB approximation are shown in Fig. \ref{nede}, with an estimate of their cumulative probability distribution, the derivative of which gives the density of states. The total fraction of atoms on the wall is therefore
\begin{equation}
	\epsilon =\int_{E_{min}}^{E_{max}} \epsilon_n \frac{dF(E)}{dE}dE=3 \pi \frac{\lambda_{T}}{R}\frac{n_{max}}{\log(E_{max}/E_{min})}\int_{E_{min}}^{E_{max}} \frac{e^{\beta E}}{E}dE
\end{equation}
and the integral is the ``exponential integral function," ${\rm Ei}(E_{max}/k_bT)-{\rm Ei}(E_{min}/k_bT)$, and using Eq. (5.1.10) of \cite{aands},
\begin{equation}\label{epsilon}
	\epsilon\approx 3 \pi \frac{\lambda_T}{R}\frac{n_{max}}{\log(E_{max}/E_{min})}\sum_{j=1}^\infty \frac{(E_{max}/k_bT)^j}{j\cdot j!}.
\end{equation}
which, for $E_{max}=.21$, $E_{min}=.0011$, with other parameters the same, yields
$\epsilon=6\times 10^{-5},$ only a factor of two larger than the previous estimate. This is because the two most strongly bound states are responsible for most of the atomic density of Hg in the surface states.  

The spin polarization relaxation rate should depend on the time that an atom resides on the surface; however, it is not clear whether $\tau_w$ is the product of a sticking time and a Boltzmann factor as has often been assumed. \cite{hap,romalis}  Given that the time between wall collisions is $\approx v/R$, which is proportional to $\sqrt{T}$, the relaxation rate should be
\begin{equation}
	\Gamma_w\propto \sqrt{T} \epsilon \propto T\sum_{j=1}^\infty \frac{(E_{max}/k_bT)^j}{j\cdot j!}\end{equation}
because $\lambda_T\propto 1/\sqrt{T}$. Noting that the fractional change in relaxation rate is readily found by taking the derivative of Eq. (\ref{epsilon}), 
\begin{equation}\label{fig4comp}
	\frac{1}{\Gamma_w}\frac{\partial \Gamma_w}{\partial T}=\frac{1}{T}-\frac{{e^{E_{max}/k_bT}}}{T}\left[
	{\sum_{j=1}^\infty \frac{(E_{max}/k_bT)^j}{j\cdot j!}}\right]^{-1},
\end{equation}
and therefore a direct comparison can be made with the data
presented in Fig. 4 of \cite{romalis}. 
The experimental values in \cite{romalis} are 0.03 at 260 K and 0.02 at 300 K, compared to the numerically determined values 0.033 and 0.024 from Eq. (\ref{fig4comp}), respectively, using $E_{max}=0.21$ from the WKB approximation.  This agreement is quite good, determined by the use of only the frequency-dependent dielectric properties of fused silica together with other measured cell parameters (temperature, distances); in some sense the calculation is ab initio.  Of course, correlations times will depend on temperature but only as the velocity. 

As an aside, it is of course possible to use the wall collision rate times the exponentially temperature dependent ``sticking time," with $\Gamma_w\propto \sqrt{T} \tau_0 e^{\beta E_0}$ in a fit to the data and approximately determine the ``surface binding energy."  The result is $E_0=0.15\ e$V which is close to $0.16\ e$V as obtained in \cite{romalis} from the same data presented in Fig. 4 of that article. Then, to get the sticking time,  a value of $\tau_0$ is required.  In \cite{romalis}, this is somewhat arbitrarily chosen as the inverse vibrational frequency of the CH$_4$-Hg van der Waals complex, $\tau_0=10^{-12}$ s \cite{vdwcomplex}.  The predicted sticking time at 293 K based on \cite{romalis} is therefore $10^{-12}\ {\rm s}\ \times e^{0.16/.025} =0.5$ ns, which is approaching an order of magnitude lower than the observed correlation times of 2 ns.

Alternatively, it is also possible to use the WKB results to estimate this frequency, by analogy with a harmonic oscillator and taking the difference in energy of the two most strongly bound vibrational states. This is about 0.02 eV, corresponding to a period $2\times 10^{-13}$ s. This could be expected because the molecular cluster vibrational frequency is determined by the reduced mass, which is essentially the CH$_4$ mass; hence, the WKB spectrum has a lower energy by $\sqrt{16/199}=0.28$, which brings the frequencies into agreement. Using the WKB results in an even larger problem, a sticking time of 0.06 ns.  It makes no sense for the correlation time to be less than the sticking time, so this approach does not really work. 

However, we do not need to consider a sticking time at all, but a dwell time. There is another time scale in the problem, which is the free flight time between wall collision, and for a 1 cm dimension cell, it is about $10^{-4}$ s.  If each traverse across the cell results in a ``sticking" collision, then the time an atom dwells in the potential well should be $10^{-4}\ {\rm s}\times \epsilon=6\times 10^{-9}$ s to maintain the statistically necessary Boltzmann distribution of atoms. The correlation times in \cite{romalis} are slightly shorter than this, which might be expected because the atoms move along the surface while in a bound state and possibly react coherently only briefly with a given relaxing site. The only requirement is that the dwell time be longer than the correlation time. 

The data as interpreted in \cite{romalis} implies the opposite, that the correlation time is longer than the sticking time, and this impossibility was left as an open question. If the probability that a wall collision results in an atom sticking is very small, the dwell time necessarily becomes longer, as required by microscopic reversibility and the value of $\epsilon$.  This implies that the approximately 6 ns dwell time obtained above is strictly a lower limit. 

The spatial extent of the surface binding region is only about 0.5 nm, so in $6\times 10^{-9}$ s, while oscillating in the potential well, an \hg atom will suffer $10^4$ collisions with the wall.  Given that there are about 100 surface states, a random walk out of the miasma of surface states could very well require $\sim 100^2$ steps, with $100^2\times 6\times 10^{-9}$ s $=6\times 10^{-5}$ s, which can be used to estimate the probability of wall collision sticking. For $10^{-4}$ s free flight time between wall collisions, the probability is $6\times 10^{-5}\ {\rm s}/10^{-4}\ {\rm s}= 0.6$.

The long wall sticking time for $^{129}$Xe as reported in \cite{xesurf} can be explained if there is only a small probability of capture in the wall potential for each wall collision, unlike the case of \hg discussed above, where it is found to be of order 60\%. Given the chemical reactivity of Hg in comparison to Xe, this is not surprising, so the sticking probability for Xe is likely smaller, and the interaction with relaxing sites might be reduced.  Both effects are needed to explain the results in \cite{xesurf}. The reported binding energy of 0.1 $e$V, determined from the variation of the relaxation rate as a function of temperature, is probably too small. In the case of \hg where, without the full theory, 0.16 eV was obtained in \cite{romalis} compared to $\approx 0.21$ eV for the most tightly bound surface obtained from the van der Waals interaction together with the WKB approximation, and the same conclusions about a ``true" larger binding energy can be drawn in the case of Xe. 

Note that $\epsilon\propto 1/R$, whereas the free flight collision time is proportional to $R$, and therefore the overall relaxation rate is independent of $R$, as expected.  We can conclude that instead of the characteristic vibration time on the wall, the relevant time scale is $\lambda_T/\bar v\propto \hbar/k_bT$.  

The possibility of using the vibrational period of an atom near the wall as a sort of sticking time is not without merit. For example, the vibrational period was used in the analysis of the relaxation of the electron spin polarization of alkali atoms as described in \cite{ambouchiat}, based on the surface physics considerations discussed in \cite{deboer}.

\subsubsection{Hydrogen surface states}

Again, using the WKB approximation, the surface states of an H atom can be determined.  The van der Waals coefficient is 2/3 that of Hg because the principal contributing optical transition for H is at a shorter wavelength. Also, the hard-sphere radius of H is smaller than that of Hg, so we can take $d_0\approx 0.12$ nm.
Together with the smaller mass of H, this leads to the result that H has only four bound states at 0.08, 0.03, 0.01, and 0.003 $e$V.  Although the energies are smaller than for Hg, $\epsilon$ is enhanced by a relatively large $\lambda_T$ compared to Hg. 
The net result for $T=293 K$, is $\epsilon=2\times 10^{-6}$, about an order of magnitude smaller than for Hg. Nevertheless, this result implies a microscopically long residence time for atomic H on the surface, during which Hg atoms can interact with the H electronic magnetic moment.

We can estimate the effective pressure of H in the bound surface states, assuming that a fraction $\epsilon$ of the total H atoms are imprisoned in a region approximately $\delta=$ 0.3 nm thick.
For a cell of radius $R$, the ratio of this surface volume to the cell volume is $3\delta/R$, so the relative densities are $\epsilon R/3\delta\approx 22$. 
For a given pressure $P$ of H gas in a 1 cm cell, the number density in the thin region near the cell wall corresponds to a pressure of about $22P$.  Alternatively, with a pressure $P_0$ in cell volume, the pressure near the surface (similar to the isothermal atmosphere problem) should be $P_0e^{E_{max}/k_bT}=P_s$. For the maximum binding energy of 0.08 $e$ V for H from the WKB approximation,
$e^{.08/.025}=24.5$ for $T=293$ K, which is in accord with the volume ratio determination.

Spin-polarization relaxation on the wall can be due to an enhanced interaction with H atoms or with the open bonds created when a surface is attacked by atomic H or excited Hg atoms. These open (covalent) bonds carry roughly an electron magnetic moment. The magnetic interaction that leads to relaxation is presented in \cite{romalis} and can be easily modified for other magnetic moments.  

\subsection{Relaxation in the Gas phase}

Paramagnetic atoms and molecules, which include oxygen and free radicals, can exist in the buffer gas as impurities or the result of a 254 nm radiation induced photochemical reaction. Reactions with an excited Hg atom can release atomic H (an unpaired spin, which is technically a free radical) that exists in the states $F=0$ and $F=1$, in proportion to 1/4 and 3/4 of total H and $F+1$ presents an effective 1/2 of an electron magnetic moment due to the addition of the proton nuclear angular momentum. 

The relaxation of nuclear spin-polarized $I=1/2$ $^1$ S$_0$ atoms by gaseous paramagnetic atoms is theoretically analyzed by Jameson, Jameson, and Hwang \cite{xeo2relax} for the specific case of O$_2$ and $^{129}$Xe,
and further developed in \cite{happerxe}, for O$_2$ on both $^3$He and
$^{129}$Xe.  Rewritten here in terms of SI units and more generally for a colliding molecule (or atom) $X$ with total angular momentum $F$ representing the sum of the internal angular momentum of the constituent nuclei and the electron spins of $X$,
\begin{equation}
	\frac{1}{T_1}=\frac{16}{3} F(F+1) ~\gamma_I^2~ \left[\mu_0~ g_F~\gamma_e\right]^2 \frac{\hbar^2}{d^2}\left(\frac{\pi \mu}{8 k_B T}\right)^{1/2}~ F(U, T)~n_0\left[X\right] 
\end{equation}
$\gamma_I$ and $\gamma_e$ are the gyromagnetic ratios for the $I=1/2$ nucleus, and for the bare electron respectively; $g_F$ is the Land\`e $g-$ factor for $X$,  and $\mu_0=4\pi\times 10^{-7}$.  $d$ is the characteristic intermolecular interaction length, $\mu$ is the reduced mass of the colliding species, $k_B$ is the Boltzmann constant and $T$ is the absolute temperature (the factor in parentheses is the reciprocal of the mean relative velocity). $n_0$ is Loshmidt constant ($n_0=2.69\times 10^{25}$/m$^3$ corresponding to the STP number density of an ideal gas), and $[X]$ is the density of the gas $X$ in Amagats.    The factor $F(U,T)$ is determined by the interaction potential $U$ between the colliding species, and is unity at high temperature in the hard-core limit. In general, it is determined by integrating solutions of the radial Schr\"odinger equation with the potential $U$, together with an average over the Boltzmann momentum distribution.  In \cite{happerxe} it is noted that the hyperfine Fermi contact term, proportional to $|\psi(0)|^4$, should generally be included in $F(U, T)$, where $\psi(0)$ is the amplitude of the electron wave function of the colliding molecule at the nucleus $I$. 
The interaction time (correlation time) for gas phase molecules is very short, so effects due to an applied magnetic field become small when the electron precession over the collision time results in a small directional change in the electron(s) spin angle(s); therefore, in all cases under consideration such effects can be ignored for the gas phase.  

In \cite{happerxe}, the experimental relaxation rate for $^3$He in O$_2$ is reported to be approximately $0.5$ /(s Amagat), which implies $F(U,T)=1.7$. For $^{129}$Xe, $F(U,T)\approx 6$ is estimated in \cite{xeo2relax}, which implies a relaxation rate of 0.7/(s Amagat). (An Amagat is the number density of an ideal gas at a pressure of 1 bar or 760 torr at 273.16 K.)

Data in \cite{pignol} indicate that the spin polarization relaxation rate of $^{199}$Hg in O$_2$ is 800/(s Amagat), which is approximately 1000 times faster than for the inert gases $^3$ He and $^{129}$ Xe.  Using the analysis for inert gases implies that $F(U,T)=1.5\times 10^4$ for \hg.  Because the two valence electrons of $^{199}$Hg enhance a hyperfine interaction between the atom and the molecule, the analysis for inert gases is not entirely accurate for Hg. Another possibility is that van der Waals molecules between \hg and O$_2$ can be formed, in which case the needed three-body collisions would make the relaxation rate pressure dependence more complicated.  This would be interesting to study.

The general conclusion is that the pressure of O$_2$ in a room temperature cell must be less than the 6 $\mu$bar level to achieve $\tau_w> 200$ s.  This implies that the oxygen impurity of the nitrogen gas used to fill the cell in Fig. \ref{200sec}  was below 10 parts per million, consistent with its spectroscopic purity grade.
Even with this high purity, the oxygen concentration in cells made using the same nitrogen had a rapid loss of Hg vapor under illumination with resonance radiation (the Hg filling pressure was 1.5 $\mu$ bars corresponding to the 0$^\circ$ vapor pressure, much lower than the oxygen pressure), as illustrated in Fig. \ref{evolution}.   

We can apply the foregoing analysis to the spin polarization relaxation of \hg by atomic H. Because H has a nuclear spin of 1/2, 3/4 of the atoms will be at hyperfine level with $F=1$ and $\gamma_s$ is reduced by a factor of two compared to O$_2$.  Comparing $F(U,T)$ with previous cases of inert gases suggests that $F(U,T)$ scales as the square root of the reduced mass of the \hg-H system, as expected from the Schr\"odinger equation. Scaling by $\mu$ and $\gamma_s^2$, and by the fraction of H atoms in $F=1$ leads to a relaxation rate for \hg nuclear spin polarization of approximately 5/(s Amagat).

A direct measure of the H-induced spin polarization relaxation rate would require the production of atomic H and a means to measure its pressure.  In light of such measurements, based on the foregoing, a total H$_2$ pressure well below a few millibars appears to be necessary to stabilize $\tau_w$ in a cell, particularly given the destructive and autocatalytic behavior of atomic H.

\section{Materials and Methods}

\begin{figure}[h]
	\centering
	\includegraphics[width=120mm]{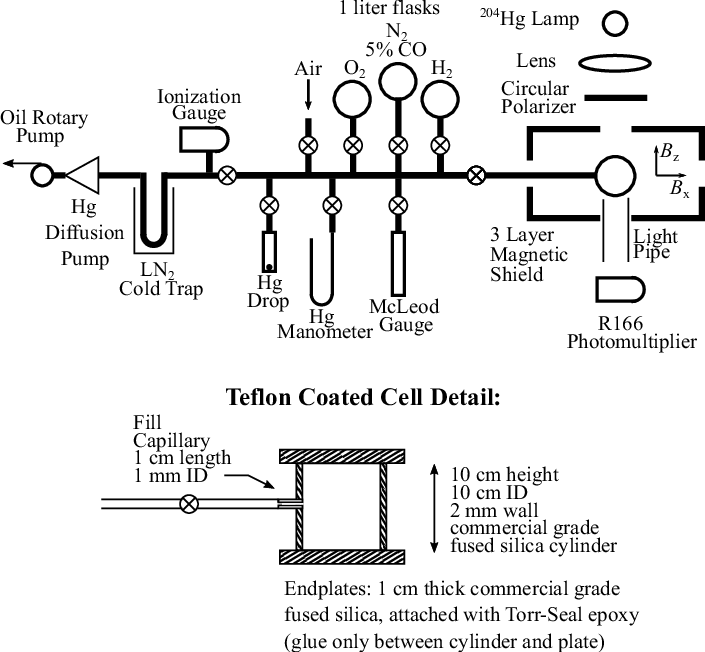}
	\caption {Manifold and optical system used for relaxation tests. See text for details.}
	\label{gassystem}
\end{figure}

To test the notions that have been discussed so far, the system shown in Fig. \ref{gassystem} was assembled in Fall 1991 at the University of Washington (these data are being reported for the first time). The system consisted of an optical pumping cell in a three-layer magnetic shield and optical components. The cell was connected to a vacuum system through a manifold that provided a selection of high purity gases (99.99\% or better)  or air that could be introduced into the cell.  The cell itself was of volume 750 m$\ell$ (we refer to these as ``1 liter cells") based on a design used to evaluate the optical pumping of \hg in larger volumes than had ever been done before, and was a key R\&D component toward the development of the system described in \cite{ksmite}.  

\begin{figure}[h]
	\centering
	\includegraphics[width=120mm]{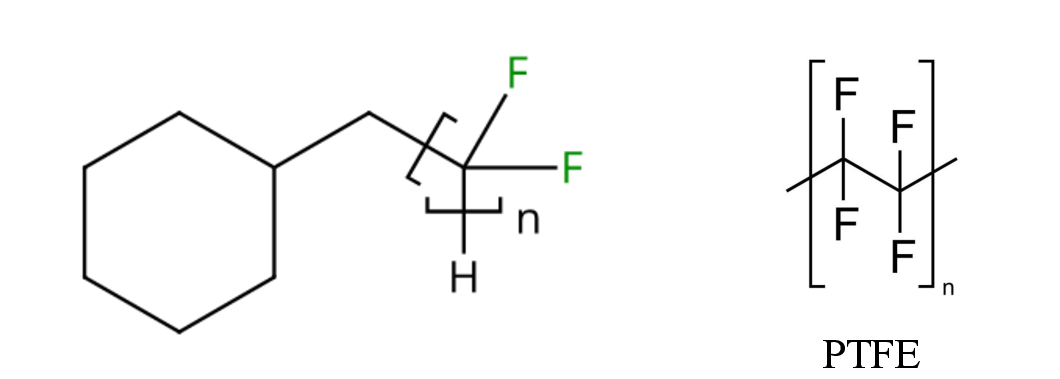}
	\caption {Comparison of a non-PTFE lubricating molecule, CAS \# 65530-85-0, Poly-tfe omega-hydro-alpha-(methylcyclohexyl) structure to the structure of pure PTFE.}
	\label{ptfestruc}
\end{figure}

The cell was coated with a generic pure PTFE lubricating spray in a hydrocarbon carrier.  Cells with a coating made using a pure PTFE lubricating spray in a primarily hydro/fluorocarbon carrier (with minor components of isopropyl alcohol and acetone, without water) had better stability than other partially fluorinated coatings or any water-based Teflon coatings.  An example is Elmer's Slide-All\rtm (which is no longer manufactured); although it was manufactured with a fast-evaporating water-free carrier, it contained a more complex polymer with hydrogen-containing functional groups at the ends of the PTFE polymer chain, as shown in Fig. \ref{ptfestruc}. These functional groups are subject to attack by excited-state Hg or atomic H and showed less stability than pure PTFE.  A lubricating spray that uses the same molecule is still available, MicroCare\rtm VDX Dry Lubricant Spray, Aerosol.

Two coats of a pure polytetrafluoroethylene (PTFE) lubricating spray were applied to the interior surface of the cylinder body, with the edges of the cylinder carefully masked. 
The end plates were masked to leave a region 1 mm wide from the outer edge of the cylinder uncoated so that the epoxy would have a region to stick between the cylinder wall and the end plate. After coating, the components were baked at 310$^\circ$ C in air and, after cooling, assembled using a minimum of epoxy and avoiding any exposure of epoxy to the inner region of the cell.    
Using Eq. \ref{mfp}, the mean free path is
\begin{equation}
	\ell=\frac{4V}{A}=\frac{4\pi R^2 h}{2 \pi R^2 + 2\pi R h}
	=\frac{2h}{1+h/R} = 6.7\ {\rm cm}
\end{equation}
where $R$ is the inner radius and $h$ is the height. 
These coatings typically gave about 20 s of $\tau_w$ per cm mfp, which implies a $\tau_w$ of
about 130 s for this large cell.  
The 1 mm fill hole, if complete depolarization occurs for \hg atoms entering that region, will lead to a relaxation rate of
\begin{equation}
	\Gamma_{hole}=\frac{1}{4} \frac{\bar v A_h}{V}\approx 40\ {\rm s}
\end{equation}
based on the atomic flux determined by kinetic theory. The observed cell $\tau_w$ was about 136 s, implying that the \hg atoms had a good chance of returning to the main volume with their polarization.  This is because the capillary inside is coated with PTFE and which, given the smoothness of the surface, indicates that Hg atoms stick briefly to the surface and re-emitted in a random direction. 

The operation of the system was straightforward.  Natural Hg was supplied to the cell from a droplet until the transmitted light was reduced by a factor of 2. Various gases could be admitted, and their pressures could be measured using a Hg manometer or McLeod gauge. A magnetic field of about 10 mG was applied along the light direction to polarize the atoms, and could be suddenly switched perpendicular to the light to observe the modulated light transmission due to the precessing atoms, as measured with a solar-blind photomultiplier.  The magnetic shields had a substantial internal magnetic gradient; these shields were constructed from scrap materials and used for testing. The gradient limited $T_2$ $\tau_w$ with gases at higher pressures; however, in those cases, the longitudinal polarization pump-up and decay could be observed, giving $T_1\approx \tau_{0}$, the wall relaxation time.  

The pumping was carried out with the full light intensity, which gave a pumping time of 60 s, while the decay was observed with the full light and also with the light intensity reduced by a factor of 10 (using an optical attenuating filter), which allowed extracting the wall decay time $\tau_w$.  

The main goal of the work was to assess the effect of hydrogen on the wall coating used in \cite{ksmite} for which $\tau_w$ degraded with the application of high voltage (HV). We surmised that the application of HV created atomic H as a result of inevitable microdischarges on the dielectric surfaces or dirt in the system.  Previous work (see Fig. \ref{200sec}) suggested that oxygen might restore the cell $\tau_w$. The secondary goal was to test for this possibility and determine how much oxygen could be used before limiting the \hg spin polarization relaxation time.  

The tests were performed with PTFE instead of deuterated polystyrene (DPS) \cite{dps} in part because in 1991 the final principal edm cell coating material used in \cite{ksmite} had not yet been decided. Typically, DPS provides $\tau_w$ of about 15 s per cm mfp, about as good as PTFE.  Resources were not available to continue the 1991 studies with DPS.

\section{Results}

The results of a series of measurements are given in Table \ref{tests}. As can be seen, there were several series of tests.  The first was to determine whether the N$_2$, 5\% CO buffer gas mix we used in \hg edm cells would affect the useful life of the cells or repair damaged cell walls.  Tests 1-5 indicate that this gas does not have a direct effect and later damage was not significantly repaired as demonstrated in Tests 12-18.

Test 6 shows that hydrogen gas alone does not damage the wall coating. 
Tests 7-12 show that Hg in the presence of resonance light immediately degrades the cell, and removal of the gas does not repair the degraded $\tau_w$. Given its stability, the degradation is probably not due entirely to the fact that atomic H is adsorbed on the surface but also to unterminated bonds on the wall coating.  Atomic H has enough energy to do this damage, as shown in Table 1, and reported in \cite{vessot,berg}. 

Tests 19-25 show that the addition of oxygen to the cell repairs the damage. We did not verify that resonance radiation is needed to make this repair (neither did we see if hydrogen gas without Hg but resonance light would cause damage).  

Tests 15 and 17 indicate that the damage might be local as the buffer gas can tend to hide damage.  The 254 nm light entered and exited the cell in a beam of about 2 cm diameter, so it is conceivable that the most significant coating damage was in these areas. 

\begin{table}[!h]
	\caption{Tests conducted with the apparatus shown in Fig. \ref{gassystem}. All $\tau_w$ measurements have an uncertainty of 8 s unless noted otherwise.}
	\label{tests}
	\centering
		\begin{tabular}{clrl}
			\toprule
			Test No. & Condition & $\tau_w$ [s] & Comment/Result\\
			\midrule
			1&only Hg vapor & 136 & Cell starting $\tau_w$\\
			2& add 1.5 torr N$_2$, 5\% CO& 133 &\\
			3& increase to 12 torr "& 60 & Gradient relaxation\\
			4& increase to 38 torr "& 30 & "\\
			5& increase to 80 torr "& 23& " \\
			6& increase to 112 torr "& 16& ", evacuate cell at end of run\\
			5& only Hg vapor & 132& cell was evacuated to base pressure\\
			6& 10 torr H$_2$& - & 10 minutes with no light, evacuate\\
			7& only Hg vapor & 132 &cell empty other than Hg\\
			8& 25 torr H$_2$& 30 & H$_2$ effect plus Gradient relaxation\\
			9& evacuate to 18 torr& 25 & "\\
			10& evacuate to 4 torr & 15$\pm 5$ & Low signal due to low Hg density\\
			11& evacuate to 0.25 torr & 25& Very low signal \\
			12& evacuate 1 hr, add Hg& 25 & Evident permanent reduction in $\tau_w$\\
			13& add 1 torr N$_2$, 5\% CO& 34 & No gradient effect expected\\
			14& increase to 50 torr "& no signal& \\
			15& increase to 110 torr "& 145$\pm 30$&pump up method so low signal;\\
			&&&improvement with buffer gas\\
			&&&implies localized damage\\
			16& evacuate, add Hg& 36& partial recovery?\\
			17& add 110 torr N$_2$, 5\%CO& $80\pm 20$ & similar to 15\\
			18& evacuate, add Hg& 36 & no apparent further recovery\\
			19& add 20 torr O$_2$& - & No observable signal\\
			20&evacuate, add Hg& 121& oxygen exposure nearly recovered $\tau_w$\\
			21& add 0.65 torr air& $\approx 7$& O$_2$ relaxation rate $\approx$840/(s Amagat)\\
			22& evacuate, add Hg &134 &fill to  high Hg density,\\
			&&& then evacuate to 1/2 light absorption\\
			23&evacuate, add Hg  & 110& evacuation overnight, to $2\times 10^{-7}$ torr \\
			24& 1 hr irradiation& 167& $\tau_w$ improving on irradiation\\
			25&no change&-&stabilized at this value\\
			\bottomrule
		\end{tabular}
\end{table}

\FloatBarrier

\section{Discussion and Conclusions}

Based on the literature on photosensitized Hg reactions, together with our own observations, the results clearly indicate that hydrogen, through atomic H, is nearly certainly responsible for much of the degradation of $\tau_w$ in \hg optically pumped magnetometry cells.  Atomic H arises from the reaction of H$_2$ with excited state Hg.  

There are two types of cells to consider: closed (permanently sealed) \hg EDM cells where the accumulation of impurities cannot be removed without breaking the cell; and open cells used in neutron EDM experiments that employ optically pumped \hg as a comagnetometer, which can be evacuated. For the latter, careful attention to the removal of hydrogenous materials, as in the case of the H-maser, which is an open cell system, should help stabilize the \hg comagnetometer system.  The addition of a small amount of oxygen to the system gave mixed results in \cite{ksmite}.   

Our 1991 measurement of the relaxation of \hg by oxygen gas agrees well with the most recent measurement reported in \cite{pignol}. Given the salubrious effect of oxygen $\tau_w$, the inclusion of a small amount will be beneficial; however, it will be consumed to form substances that stick to the wall or mix with wall coating materials. In addition, Hg will be consumed to form HgO, so a source of \hg within the cell is needed, or ozone production must be suppressed.  Given that of order $10^{13}$ photon per second are incident on a cell and that the photosensitized reaction in some cases has a quantum efficiency near unity, we might expect the same number of atoms or molecules lost per second. This corresponds to $10^{18}$ reaction per day, or 10 mbar of gas consumed per day in a 5 m$\ell$ cell.

According to Table 1, the only (single) bond that is immune to damage is HF.  In addition, the energy available from the Hg bonding to H is available to break other bonds if the vehicle of destruction is HgH.  

The addition of CO greatly improved the longevity of cell usefulness, and anecdotal evidence suggests that scrupulous cleanliness and the avoidance of hydrogen gave further improvement, however, the fabrication of cells remains hit-or-miss.  A principle effect of CO is to quench the metastable excited state of Hg, which carries enough energy to break bonds.

Hydrogen likely does its damage through an autocatalytic reaction, in that if there is any hydrogen, it will lead to photosensitized production of more hydrogen by breaking the bonds of hydrogenous materials on the cell walls.

It was hoped that CO might bind to atomic H and eventually form formaldehyde, which would get buried in the wall coating. However, the formyl intermediary is a free radical that probably causes relaxation at a rate similar to that of oxygen, and the hydrogen in both HCO and H$_2$CO are sources for photosensitized hydrogen production.

Identifying the surface states allows for a theoretically consisted picture to explain the data presented in \cite{romalis}.  In all, this was quite remarkable and unexpected.

Similarly, the seemingly anonymously long sticking times for $^{129}$Xe as reported in \cite{xesurf} can be explained through surface states together with a low probability of sticking per wall collision, which is expected for Xe compared to Hg.  Using microscopic reversibility and the calculable value of $\epsilon$ as the average fraction of atoms stuck in the wall potential, the probability of sticking can be determined, which is of the order 1\%.  

Rather than the sticking time, there is another time scale that is better suited for the case of heavy atoms with many bound states. That time scale is the thermal wavelength divided by the average velocity, which is proportional $\hbar$ divided the kinetic energy of the atom. This time scale is the inverse of the ``thermal frequency,'' in analogy to the thermal wavelength. 

The large number of relatively deep surface states (compared to $k_bT=0.025$ eV) brings into question the efficacy of \hg co-magnetometry.  For an extremely precise magnetometry system, as has been proposed, the surface dwell time (during which a substantial spin polarization can survive) will need to be more carefully considered, as the dwell time can lead to corrections, e.g., localized systematically generated magnetic fields are sensitive to the volume averaging. The gravitation offset of ultracold neutrons in neutron EDM experiments is perhaps a more serious problem; however, all such effects will eventually need consideration. 

Based on the foregoing, to improve the durability of closed or open \hg optical pumping and detection cells (i.e. any similar cells used for fundamental experiments), some guidelines are as follows:
\begin{enumerate}
	\item Avoid hydrogen gas either as an impurity in the CO buffer gas (oxygen can be removed, but water as an impurity will produce hydrogen which will pass through the purification amalgam) or generated by photosensitized reactions. Sealing of filled fused silica cells involves much heat and even UV light is produced, so care is needed in such operations;
	\item For open cells used in a neutron EDM experiment that employs a \hg comagnetometer, the problem of $\tau_w$ degradation appears to be largely solved, in particular by use of an oxygen gas rinse. \cite{psiedm,psi}  The use of deuterated polystyrene offers some stability from attack by atomic H and other free radicals. 
	This is in part due to the 1.6 eV energy required to disrupt the resonance stabilization of the phenyl ring. With this additional energy, an H atom cannot cause a C-H bond to break to form H-H; however, atomic H can be added to the phenyl ring, thus producing a cyclohexadienyl radical,\cite{ahdps} while exposure to ozone and oxygen leads to cross-linkages in the polymer.\cite{o3dps}  This additional energy presents a barrier to direct reactions between benzene molecules and excited-state Hg atoms, as can be seen from the energies in Table 1.  A study of the reaction between atomic O and benzene used the photosensitized reaction between Hg and N$_2$O to produce O in situ in a reaction vessel, with the final amount of nitrogen as a measure of the total amount of reacted oxygen.\cite{c6o}  
	
	Pertaining to high voltage stability, the phenyl ring is attacked by positive ions that can bind to the electron cloud of the ring, leading to destabilization of the molecule due to disturbance of the resonance of the ring. This implies that DPS-coated cells require that high voltage can be applied without electrical discharges that produce positive ions. This also implies that DPS is less stable under high voltages than aliphatic compounds. 
	
	\item Photosensitized reactions in closed cells can be suppressed by use of a quenching gas; CO has proven effective but there are better gases;
	\item N$_2$O is among the best gas for quenching both \ptrip and \psing excited Hg atoms\cite{japanquench}.  
	\item N$_2$O can be broken by photosensitized reactions with Hg, as is possible with the bond energies presented in Table 1.  However, ozone formation, necessary to form HgO, is suppressed due to a scavenging reaction with N$_2$O.  Preliminary studies indicated that it might be possible to produce a cell with N$_2$O in a proportion with an inert gas, perhaps a mixture of CO and Xe, that will produce oxygen at a rate sufficient to maintain a long stable life, for example, similar to the data present in Fig. \ref{200sec}, but without the loss of Hg due to the formation of HgO.  
	\item Use fluorinated waxes or silanes with less hydrogen, e.g., Trichloro(3,3,3-trifluoropropyl)silane (CAS 592-09-6). A fully fluorinated silane would be of interest; however, none appears to be commercially available. Tests with fluorinated wax have not shown significant improvement but appear as promising.\cite{lindpc,fwax}
	\item Deuterated polystyrene (DPS) \cite{dps} remains one of the few options for UCN \hg comagnetometer experiments.  
\end{enumerate}

\section*{Acknowledgments}

The author thanks Prof. L.R. Hunter of Amherst College for reporting on the behavior of his cells, Eric Lindahl, the University of Washington glassblower, for technical advice regarding material preparation and for quartzware fabrication, and Jennie Chen for discussions of magnetometry in connection to the LANL EDM project. A comment on the manuscript by Prajwal Mohan Murthy confirming the improvement of $\tau_w$ for \hg after an oxygen flush is appreciated.  I began this work under the direction of Prof. E.N. Fortson in 1982 who was my Ph.D. advisor and Prof. B.R. Heckel, both of the University of Washington, Seattle, WA.  Discussion with Prof. J.M. Pendlebury (deceased) of Sussex University led to the recognition of surface states and their role in the behavior of optical pumping cells.

\section*{Funding Statement}
This research was sponsored by Yale University.
\section*{Conflict of Interest}
The author declares that there is no conflict of interest.
\section*{Institutional Review Board Statement}
Not applicable.

\section*{Data Availability}

All data used in this study are presented in this report or are fully contained in the cited literature.

\bigskip
S.K.L. is solely responsible for the conception, research, writing, and finalization of this article. The author will read and agree to the published version of the manuscript.

\bibliographystyle{vancouver}
\bibliography{HgH}

\end{document}